\documentstyle[epsfig]{EuroPhys}
\input EuroMacr
\begin{document}

\Date{28 April 1997}

\title{
Ground State Laser Cooling Beyond the Lamb--Dicke Limit
}

\author{G. Morigi\inst{1}, J.I. Cirac\inst{1}, M. Lewenstein\inst{2}, 
\And P. Zoller\inst{1}}

\institute{
\inst{1} Institut f\"ur Theoretische Physik, Universit\"at Innsbruck,
Technikerstrasse 25, A--6020 Innsbruck, Austria.\\
\inst{2} CEA/DRECAM/SPAM, Centre d'Etudes de Saclay, 91191 Gif--sur--Yvette Cedex,
France
}

\pacs{
\Pacs{32}{80Pj}{Optical cooling of atoms; trapping.}
\Pacs{42}{50Vk}{Mechanical effects of light on atoms.}
}

\maketitle
\begin{abstract}
We propose a laser cooling scheme that allows to cool a single atom
confined in a harmonic potential to the trap ground state $|0\rangle$.
The scheme assumes strong confinement, where the oscillation frequency
in the trap is larger than the effective spontaneous decay width, but is
not restricted to the Lamb--Dicke limit, {\it i.e.} the size of the trap
ground state can be larger than the optical wavelength. This cooling
scheme may be useful in the context of quantum computations with ions
and Bose--Einstein condensation.
\end{abstract}


One of the major goals of atomic physics is to laser cool trapped atoms
to the lowest energy state $|0\rangle$ of the confining potential. While
this is certainly possible in the so--called Lamb--Dicke Limit (LDL)
\cite{Di89}, whereby the spatial dimensions of the ground state
$|0\rangle$, $a_0$, are much smaller than the wavelength of the cooling
laser $\lambda$ ({\it i.e.} $\eta=2\pi a_0/\lambda \ll 1$)
\cite{reviews,review3}, there is not known mechanism that achieves this
task in the opposite limit ($\eta\geq 1$) \cite{reviews2}. Ground state
cooling beyond the LDL is interesting from two perspectives. On the one
hand, it is required to perform quantum computations in a linear ion
trap \cite{Ci95,Mo95}, as well as to produce some non--classical states
of motion of a single ion or atom in scales of the order of the laser
wavelength \cite{review3}. On the other hand, it may be a way to obtain
Bose--Einstein condensation \cite{BECwieman,BECkett} with all--optical
means \cite{reviews2,review4}. 

So far, the most efficient laser cooling mechanism for trapped particles
in the LDL is {\em sideband cooling} \cite{Neuhauser}. It allows to cool
a single particle confined in a harmonic trap practically to the ground
state, as first demonstrated experimentally by Diedrich {\it et al.}
\cite{Di89}. Beyond the LDL, the most effective techniques have been
proved to be cooling schemes originally designed to achieve subrecoil
temperatures of free atoms, namely, dark state \cite{Dark} and Raman
cooling \cite{Raman}. The first scheme operates with an angular momentum
$J_g=1\rightarrow J_e=1$ internal transition, and it is specially
suitable for flat--bottom traps \cite{Pellizari}, whereas the second one
is based on timing laser pulses followed by a repumping process,
creating and populating a dark state of zero velocity atoms. In this
paper we propose a method which extends the 
sideband cooling mechanism beyond the LDL, combining ideas of these other 
schemes. It allows to cool a single atom in a harmonic trap to the ground 
state of the confining potential for $\eta \geq 1$. Although we will present
our scheme in a one dimensional situation, it can be 
easily generalized to higher dimensions, choosing appropriately the geometry of
the laser beams; furthermore, it can be easily adapted to the case in which one has many ions in an 
electromagnetic trap, as in the ion trap quantum computer 
model~\cite{Ci95,Mo95} by applying it to the cooling of the center-of-mass
mode. Note that ground state cooling is a necessary condition
to perform quantum gates with ions. In addition, for a set of neutral atoms it 
operates in the regime in which the heating mechanisms due to reabsorptions 
can be minimized~\cite{Europh}, and therefore it is a promising alternative 
to evaporative cooling to achieve Bose--Einstein condensation.

Standard sideband cooling of a two-level atom confined in a one--dimensional
harmonic trap is achieved by tuning a laser beam to the lower motional
sideband of an internal transition ({\it i.e.} $\delta=-\nu$,
where $\delta$ is the laser detuning from the two--level resonance, and
$\nu$ is the trap frequency). Denoting by $|n\rangle$ ($n=0,1,\ldots$)
the eigenstates (Fock states) of the harmonic potential, and by
$|g\rangle$ and $|e\rangle$ the two internal atomic states, sideband
cooling can be summarized as follows: 

{\bf (i)} The laser induces transitions $|g,n+1\rangle \rightarrow
|e,n\rangle$. In this way, if spontaneous emission mainly gives rise to
the $|e,n\rangle \rightarrow |g,n\rangle$ transition, in each
absorption--emission cycle a quantum of motional energy $\hbar \nu$ is
removed. This is precisely the case in the {\it Lamb--Dicke limit}, since
$\eta^2= \epsilon_R/(\hbar\nu) \ll 1$ ($\epsilon_R$ is the photon
recoil), and therefore the kinetic energy provided by the emitted photon
is not sufficient to excite a vibrational quantum. 

{\bf (ii)} At the end of the process, the particle remains in the state
$|g,0\rangle$, where it can no longer be excited by the laser; thus, the
state $|g,0\rangle$ is a ``dark state". This last condition requires to
operate in the {\it strong confinement limit}, whereby the bandwidth of
the excited internal state $\Gamma$ is smaller than the trap frequency
($\Gamma < \nu$).

\begin{figure}
\begin{center}
\epsfig{file=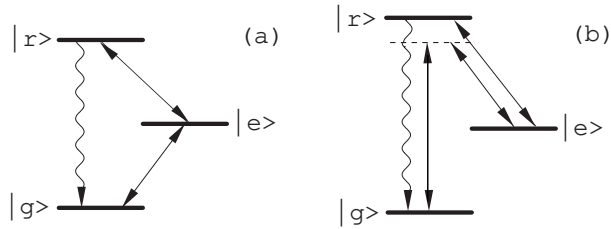,width=8cm}
\caption{
Internal configuration of a three level atom, that can be reduced to an
effective two level one with ground state $|g\rangle$ and excited state
$|e\rangle$. In (a) $|g\rangle\rightarrow |r\rangle$ 
is an electric dipole allowed transition, 
whereas $|g\rangle\rightarrow |e\rangle$ and $|e\rangle\rightarrow |r\rangle$
are electric dipole forbidden ones; 
in (b) $|g\rangle\rightarrow |r\rangle$ $|r\rangle\rightarrow |e\rangle$ are 
electric dipole allowed transition. A laser on resonance on the transition 
$|e\rangle\rightarrow |r\rangle$ pumps optically the atoms into $|g\rangle$.
}
\label{fig1}
\end{center}
\end{figure}

Despite the fact that for optical dipole transitions typically $\Gamma >
\nu$, one can satisfy the strong confinement condition by either using a
metastable excited state and quenching its population using a third
(fast decaying) level [Fig.~1(a)], or by using a Raman transition
[Fig.~1(b)]. In both cases, given that the extra level is always almost
empty, one can eliminate it adiabatically, and obtain an effective
two--level transition with a ``designed'' effective spontaneous rate
$\Gamma_{\rm eff} < \nu$ \cite{Marzoli}. The condition related to the
LDL, however, cannot be manipulated in a similar way. For $\eta\geq 1$,
spontaneous emission will spread the population of the level
$|e,n\rangle$ into $|g,n\pm m\rangle$, where $m\sim 0,1,\ldots,{\cal
O}(\eta^2)$, since the spontaneously emitted photon will typically
provide the atom with one recoil $\epsilon_R = \eta^2 \hbar \nu$ of
energy. This will produce a diffusing (heating) mechanism that will
prevent most of the atoms from ending up in the ground state $|g,0\rangle$.

Our proposal overcomes the problem of diffusion, leading to ground state
cooling. It consists of driving the atom(s) with two sequences of laser
pulses:

{\bf (1)} The {\it first} pulse plays a {\it confinement} role. It
drives the atomic population to the levels $|g,n\rangle$ with energy
less than one recoil ({\it i.e.} $n \leq \eta^2$).  This is achieved
by selecting the laser detunings $\delta \simeq -\hat{\eta}^2 \nu$, where
$\hat \eta^2$ denotes the closest integer $\ge \eta^2$. The effect of
these pulses is similar to the traditional sideband cooling: (i) the
laser induces transitions $|g,n+{\cal O}(\eta^2)\rangle \rightarrow
|e,n\rangle$, whereas spontaneous emission produces transitions
$|e,n\rangle \rightarrow |g,n+m\rangle$, with $-\eta^2 \leq m
\leq\eta^2$, and therefore some energy is in average lost in each
absorption--emission cycle; (ii) At the end of the pulse sequence, all
the population remains in the states $|g,n\rangle$ with $n\leq
\eta^2$, since they are dark with respect to the laser.

{\bf (2)} The second pulse sequence pumps atoms into the ground state.
This is achieved by emptying the motional levels ($|g,1\rangle$,
$|g,2\rangle$, $\ldots$) while leaving the population of the ground
level $|g,0\rangle$ basically untouched. This purpose can be realized
{\it (i)} through an appropriate selection of the pulses frequency. In
each cycle, some of these atoms will fall in the ground level
$|g,0\rangle$, while the ones heated will be then reconfined after a
repetition of the first sequence of pulses. An important ingredient of
this scheme is that beyond the LDL, these lasers have to be {\it blue
detuned}. {\it (ii)} Another realization of the emptying stage can be
achieved through the application of Rabi pulses which are multiples of
$2\pi$ cycles for the ground state $|g,0\rangle$, so that
$|g,0\rangle$ is a trapping state~\cite{trapping}. This requires that
the pulses are sufficiently short so that incoherent processes can be
neglected.  From now on we will discuss the {\it (i)} emptying scheme,
for which a rate equation treatment is suitable.

Let us start considering a two level internal transition.
Assuming $\Omega\ll\Gamma$, where $\Omega$ is the Rabi frequency and $\Gamma$
the excited state decay rate, and $\nu\gg \Omega^2/\Gamma$, one can
eliminate the internal excited level, and obtain a set of rate
equations for the populations $P^g_n=\langle g,n|\rho|g,n\rangle$:
\begin{equation} 
\label{rateeq}
\dot P_n^g = -\left[\sum_{m=0}^\infty
\Gamma_{m\leftarrow n}\right] P_n^g + \sum_{m=0}^\infty
\Gamma_{n\leftarrow m} P_m^g. 
\end{equation} 
According to Eq.~(\ref{rateeq}) the population
of level $|g,n\rangle$, increases (decreases) due to transitions from 
(to) the levels $|g,m\rangle$ at a rate $\Gamma_{n\leftarrow m}$ 
($\Gamma_{m\leftarrow n}$). This rate is given by
\begin{equation}
\label{Rate1}
\Gamma_{n\leftarrow m} = \frac{\Omega^2}{\Gamma}
\int_{-1}^1 du N(u) \\
\left| \sum_{k=0}^\infty 
\frac{\gamma \langle n| e^{-ik_Lxu} |k\rangle
\langle k|e^{ik_Lx}|m\rangle}
{\delta-\nu (k-m) + i\gamma} \right|^2, 
\end{equation}
where $k_L$ is the laser-field wavevector and $N(u)$ angular distribution 
of the spontaneously emitted photon.
The rate at which level $|n\rangle$ is emptied is given by 
\begin{equation}
\Gamma_n = \sum_{m\ne n} \Gamma_{m\leftarrow n} \simeq
\frac{\Omega^2}{\Gamma}\sum_{k}
\frac{\gamma^2 \left| \langle k|e^{ik_Lx}|n\rangle \right|^2}
{[\delta-\nu(k-n)]^2+\gamma^2}.
\end{equation}
In the strong confinement limit ($\gamma \ll \nu$), and for a laser
frequency such that $\delta = -\nu k_0$ (for certain integer number
$k_0$), we can neglect the non--resonant terms to obtain 
\begin{equation}
\label{rateout}
\Gamma_n \simeq \left\{ 
\begin{array}{ll}
\frac{\Omega^2}{\Gamma}\left|\langle n-k_0| e^{ik_Lx}|n\rangle
\right|^2 & \mbox{ if $n>k_0$},\\ 
        0 & \mbox{ otherwise}.
\end{array}
\right.
\end{equation} 
%

\begin{figure}
\begin{center}
\epsfig{file=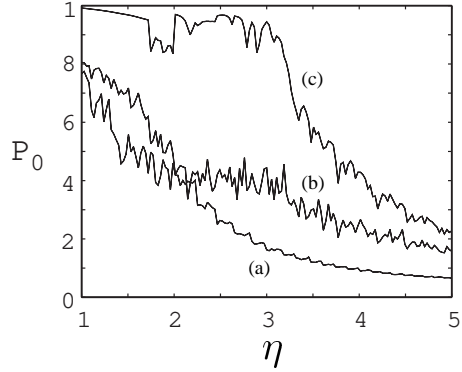,width=6cm}
\caption{
Population in $|g,0\rangle$ as a function of $\eta$ after 200 sequences of pulses,
for $\Gamma=0.1\nu$, and each pulse of duration $t=0.1\nu^{-1}$. 
The curve (a) represents a sequence of one single pulse detuned
with $\delta=-{\rm max}(2,(\hat \eta^2))\nu$; (b) represents a sequence of one 
single pulse with $\delta=-\nu$; (c) represents a sequence of pulses 
respectively with $\delta_1=-{\rm max}(2,(\hat \eta^2))\nu$ and 
$\delta_2=-\nu$.  
}
\label{fig2}
\end{center}
\end{figure}

In Fig.~2 we have plotted the population $P_0$ as a function of the
Lamb-Dicke parameter $\eta$, after 200 cycles for different
detunings. The curves have been obtained by solving numerically the
rate equations (\ref{rateeq}) with a truncated set of states, using
the rates (\ref{Rate1}), where the initial distribution was taken to
be a thermal one with mean $\langle n\rangle =6$. Curves (a) and (b)
correspond to cycles with a single laser pulse of detuning
$\delta_{a,b}=-\nu$ and $- {\rm max}(2,\hat \eta^2)\nu$, respectively,
where $\hat \eta^2$ denotes the closest integer $\ge \eta^2$. Curve
(c) on the other hand corresponds to cycles composed of two different
laser pulses of detunings $\delta_{a,b}$, respectively. As this figure
shows, with this last scheme one has more than $90\%$ of the atoms in
the ground state, even for value of $\eta$ up to $3$. In fact,
according to (\ref{rateout}), the first laser pulse confines the atoms
in the levels $|n\rangle$ with $n<\eta^2$, which are approximately
dark with respect to this laser configuration \cite{footnote},
provided that the pulse durations are $t\sim \Gamma_{n\approx{\hat
\eta}^2}$. Furthermore, the population in $|n\rangle$ with $n\ge{\hat
\eta}^2$ will be distributed among states $|m\rangle$ with $n-2\eta^2
\leq m \leq n$ in each laser absorption--spontaneous emission cycle
\cite{note1}. The accumulation of atoms in $|g,0\rangle$ is achieved
by using a laser pulse tuned to the lower motional sideband, as in
standard sideband cooling.

For $\eta\ge 3$, however, the efficiency of the method drops
drastically. The reason for this behaviour lies in the form of the
Franck-Condon coefficient appearing in (\ref{rateout}), since it
becomes exponentially small for large $\eta$ and $k_0\simeq
0$. Therefore, the laser pulse tuned to the lower motional sideband
loses its effectiveness, and one has to find another way to empty all
the levels different from $|g,0\rangle$. This can be achieved by using
{\em blue detuned} pulses, provided they are chosen in such a way that
$\langle k_0|e^{ik_Lx} |0\rangle$ ({\it i.e.} $\Gamma_0$) is small,
but $\langle k_0+1|e^{ik_Lx} |1\rangle$ ({\it i.e.} $\Gamma_1$) is
sufficiently large, as explained below. We show in Fig.~3 
the final populations $P_n$ after 200 cycles of laser pulses, for $\eta=5$. 
In Fig.~3(a) we
have simulated numerically cycles with a single laser pulse of
duration $t_1=0.6 \Gamma/\Omega^2$ and detuning $\delta_1=-24\nu$
({\it i.e.} $\delta_1\simeq - {\cal O}(\eta^2) \nu$). In Fig.~3(b) we
have simulated cycles consisting of 4 laser pulses of durations
$t_{1,2}=0.6 \Gamma/\Omega^2$, $t_{3,4}=0.2 \Gamma/\Omega^2$, and
detunings $\delta_{1,2,3,4}=-24\nu,-25\nu,+7\nu,+9 \nu$, respectively. As the
figure shows, in this last case one can obtain about a 90\% of the
population in the ground state of the harmonic potential. Note that by
choosing a blue detuned pulse with $\delta=+7\nu$ and a pulse duration
such that $\Gamma_0 t \ll 1$ the excitation of the state $|g,0\rangle$
is negligible. This pulse, however still empties states $|g,n\rangle$
($n\ge 1$), since $\Gamma_0/\Gamma_1 \ll 1$ as it is displayed in
Fig.~4, where the rates $\Gamma_n$ are plotted in function of $n$. 
Note that, due to the
oscillatory nature of the Franck--Condon coefficient, some states with
$n\ne 0$ can be negligibly coupled to the laser field during both the
confining pulses and the emptying pulses, leading to a loss of
effectiveness of the cooling mechanism. This problem can be overcome
by applying, while confining the atoms, a second laser pulse with
$\delta=-(1+\eta^2)\nu$, and while emptying a level, another blue
detuned pulse that satisfies the requirements cited above. This is
precisely the role of the pulses with detunings $\delta_{2,4}$ of
Fig.~3. Finally, it is worth mentioning that the cooling time scale is
independent of the initial temperature of the atoms, when the
condition $\langle n\rangle\leq {\cal O}(\eta^2)$ is fulfilled, {\it
i.e.} when the initial atoms temperature is below the recoil limit.

\begin{figure}
\begin{center}
\epsfig{file=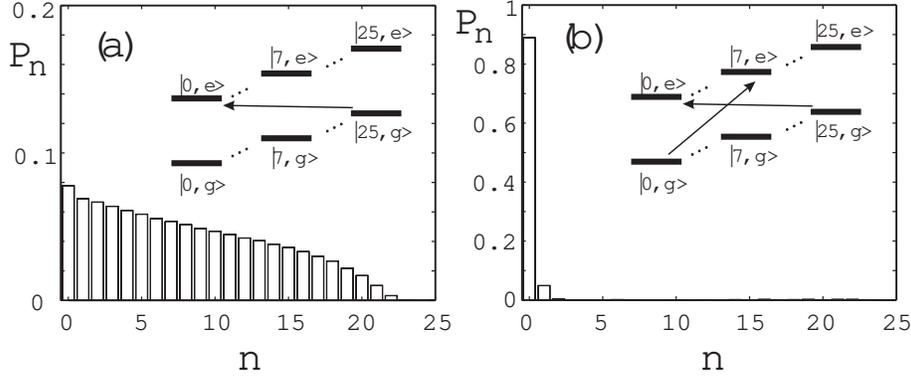,width=12cm}
\caption{
Probability distributions $P_n$ for $\Gamma=0.1\nu$ and
$\eta=5$ after 200 cycles, starting from a thermal state with 
a mean number 6. Each cycle is composed of:
(a) single laser pulse of duration $t_1=0.6 (\Gamma/\Omega^2)$ and
detuning $\delta_1=-24 \nu$; (b) four laser pulses of duration
$t_1=t_2=0.6 (\Gamma/\Omega^2)$ and $t_3=t_4=0.2 (\Gamma/\Omega^2)$,
and detunings $\delta_{1,2,3,4}=-24\nu, -25\nu, 7\nu$, and $5 \nu$.
}
\label{fig3}
\end{center}
\end{figure}

\begin{figure}
\begin{center}
\epsfig{file=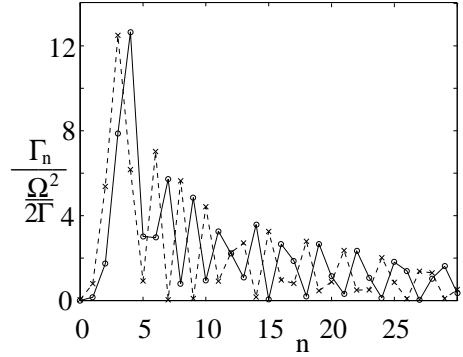,width=6cm}
\caption{
$\Gamma_n$ (in units of $\Omega^2/\Gamma$) as a 
function of the Fock number $n$, for $\eta=5$. The
curve with `$\circ$` corresponds to $\delta=7\nu$ and the one with `$\times$` 
to $\delta=9\nu$. In both cases the ratio $\Gamma_0/\Gamma_1\simeq 0.04$.
}
\label{fig4}
\end{center}
\end{figure}

Calculations performed for the cooling scheme based on trapping states
yield very similar results where typically more than $90\%$ of the
atoms are pumped to the ground state.

In most of the practical situations, the strong confinement condition
$\Gamma < \nu$ has to be achieved either by using a Raman transition or
by quenching a metastable level [see Figs.~1(a,b)], as in laser cooling
in the LDL. On the other hand, we have used here a 1--dimensional model.
Whereas this is the situation in a linear ion trap, it is not so in
Bose--Einstein condensation experiments. In that case, one should
alternate pulses along different directions, as one does in dark state
cooling \cite{Dark}. 

In conclusion, we have proposed here a novel laser cooling scheme for
trapped particles that operates in the strong confinement limit and
is not restricted to the LDL, that makes possible the cooling of the atoms
into the ground state of the harmonic confining potential. This scheme may
be useful in quantum computing as well as Bose--Einstein condensation.\\ \\

\stars
We thank R.~Blatt, D.~Wineland and
the ion trap group at the Max-Planck Institute for Quantum Optics in Munich 
for discussions. Work at Innsbruck
is supported by the Austrian Science Foundation. Work supported by TMR
networks ERBFMRX-CT96-0087 and ERBFMRX-CT96-0002.



\begin{thebibliography}{99}
\bibitem{note1}
By selecting $k_0<\eta^2$, some of the population will diffuse to
states with $m>n$ thus yielding heating. For $k_0 > \eta^2$, 
the population in some levels with an energy larger than the recoil
$\epsilon_R$ will not decrease.
\end{thebibliography}
\end{document}